\documentclass[10pt,a4paper]{article}

\usepackage{graphicx}
\usepackage{amssymb,amsmath}
\usepackage{a4wide}

\newtheorem{thm}{Theorem}
\newtheorem{remark}{Remark}
\begin{document}

\title{On the Capacity of the Single Source Multiple Relay Single Destination Mesh Network}

\author{Lawrence Ong and Mehul Motani\\
Department of Electrical and Computer Engineering\\
National University of Singapore\\
Email: \{lawrence.ong, motani\}@nus.edu.sg}
\date{}

\maketitle

\begin{abstract}
% Text of abstract
In this paper, we derive the information theoretic capacity of a special class of mesh networks. A mesh network is a heterogeneous wireless network in which the transmission among power limited nodes is assisted by powerful relays, which use the same wireless medium. We investigate the mesh network when there is one source, one destination, and multiple relays, which we call the single source multiple relay single destination (SSMRSD) mesh network. We derive the asymptotic capacity of the SSMRSD mesh network when the relay powers grow to infinity. Our approach is as follows. We first look at an upper bound on the information theoretic capacity of these networks in a Gaussian setting. We then show that this bound is achievable asymptotically using the compress-and-forward strategy for the multiple relay channel. 
We also perform numerical computations for the case when the relays have finite powers.
%While this result is derived for the asymptotic case when the relays' powers grow to infinity, we extend our analysis to the case with finite but large relay's power (using the single relay network as an example). 
We observe that even when the relay power is only a few times larger than the source power, the compress-and-forward rate gets close to the capacity. The results indicate the value of cooperation in wireless mesh networks. The capacity characterization quantifies how the relays can cooperate, using the compress-and-forward strategy, to either conserve node energy or to increase transmission rate.
\end{abstract}

% main text
% \appendix

% \section{}
% \label{}

\section{Introduction}
Wireless networks have been finding more applications and capturing much research attention in recent years. The advantage of mobile clients makes ad-hoc wireless networking an attractive solution for home and enterprise users. However, with an almost unlimited number of ways of interacting and cooperating, analysis of these multi-terminal networks is difficult. To date, the capacity of even the simple three-node channel~\cite{meulen71} is not known, except for special cases, for example, the multiple access channel \cite{liao72}\cite{ahlswede74}, the degraded relay channel \cite{covergamal79}, the degraded broadcast channel \cite{bergmans73}. However, this did not hinder research in channels with more nodes.

\begin{figure}[t]
\centering
\includegraphics[width=10cm]{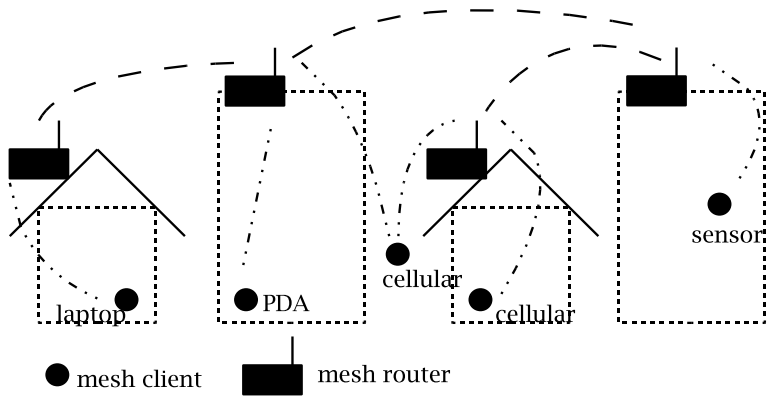}
\caption{A mesh network.}  \label{fig:mesh}
\end{figure}

Recently, \emph{mesh networks} have been drawing interest from both the research arena and the industry. Mesh networks (\cite{akyildiz05} and the references therein) are peer-to-peer multihop wireless networks with powerful relays. One practical setup of the mesh network is depicted in Fig.~\ref{fig:mesh}. It consists of mesh routers (stationary and line powered) and mesh clients (mobile and battery powered). This type of setup is adopted by 802.11s \cite{hauserbakerconner04}, the IEEE standard (waiting for approval) for wireless mesh networks. The standard defines a network of wireless access points (mesh routers) communicating wirelessly with each other and serving mesh clients in their proximity. In other words, mesh routers act as relays. The access points forward data packets via multi-hopping. They do not need to connect to the wired backbone.

%The mesh network is able to self-configure and its distributed nature makes it robust to single router failure. 

In this paper, we study the information theoretic capacity of a class of mesh networks, which include but are not restricted to the models defined by IEEE 802.11s. We study the single source multiple relay single destination (SSMRSD) mesh network, which we model by the multiple relay channel \cite{xiekumar03}\cite{kramergastpar04}\cite{ongmotani05a}\cite{ongmotani05b}\cite{chongmotani05b}. The multiple relay channel captures the scenario where the transmission from the source to the destination is aided by several relay nodes, which themselves have no data to send. As such, the SSMRSD mesh network is a multiple relay channel where the relays can transmit at large power. One can treat the SSMRSD mesh network as an excerpt of the general mesh network where we consider just one of the source-destination pairs. We note that the capacity of the general multiple relay channel has not been found, except for special cases, e.g., the degraded multiple relay channel.  The capacity of the SSMRSD mesh network, which is not a degraded multiple relay channel~\cite[Theorem 2.3]{xiekumar03}, has not been found.

We investigate rates achievable by the compress-and-forward coding strategy on the SSMRSD mesh network. The coding technique was first introduced by Cover and El Gamal \cite{covergamal79} for the single relay channel and later extended to the multiple relay channel by Kramer \emph{et al.} \cite{kramergastpar04}. In the compress-and-forward strategy, the source transmits to both the relays and the destination. The relays do not decode the data, but simply quantize, compress, and send them to the destination. We categorize the compress-and-forward strategy as a cooperative coding strategy. To understand why, we contrast it with a non-cooperative coding strategy, the \emph{multihop strategy}. In the multihop strategy, when one node transmits, only the next hop listens. Though other nodes over-hear the transmission, they regard it as noise. On the other hand, in the cooperative strategy, instead of ignoring the useful transmission, nodes that over-hear the transmissions of other nodes (though the information might not be intended for them) help to forward the data, fully or partially, to the destination.

We show that the compress-and-forward strategy approaches the capacity of the SSMRSD mesh network as the relay powers grow. Our approach is as follows. First, we study an upper bound on the capacity of the SSMRSD mesh network, which is derived from a max-flow min-cut argument. We then study achievable rates of the compress-and-forward coding strategy on the multiple relay channel.  We show that when the power constraints at  the relays are relaxed (which is the case in the SSMRSD mesh network), the compress-and-forward technique approaches the capacity upper bound asymptotically. Using the single relay channel as an example, we numerically investigate the performance of the compress-and-forward strategy as the relay power grows.

%What our results say for mesh network design is that cooperative relaying (via decode-and-forward or compress-and-forward) by mesh routers can facilitate communication between mesh clients.  
%Recent work \cite{razaghiyu06} indicates how one can design practical coding schemes to implement this cooperative relaying.  
%Our results on the effects of relay's position can help clients to determine which mesh routers to use and what kind of cooperation to employ.  
%Combining this with network layer algorithms will lead to an efficient protocol stack for mesh networks, on which a variety of user applications can be built.

Our contributions in this paper are as follows:
\begin{enumerate}
\item We derive an achievable rate expression for the compress-and-forward strategy on the Gaussian multiple relay channel.
\item We derive the capacity asymptotically of the SSMRSD Gaussian mesh network as the relay powers grow to infinity.
\item We show, using numerical computations, how the compress-and-forward rate approaches the capacity as the relay power grows. More specifically, we show that the rate achievable by the compress-and-forward strategy is close to the capacity even when the relay power is only a few times higher than the source power.
\item We study and compare coding strategies on the SSMRSD mesh network. The comparison of different coding strategies motivates code designers to design practical coding schemes based on certain types of cooperation, to drive the transmission rate toward the theoretical limits. Knowing the network capacity as a function of transmission powers, and understanding how nodes should cooperate help network designers in network deployment.
\end{enumerate}

%Analysis on the capacity of multiple source multiple destination networks is hard. Gupta and Kumar \cite{guptakumar00} found an upper bound and an lower bound on the capacity of such channels. However, the analyses cannot be readily applied to mesh networks. This is because in \cite{guptakumar00}, every node has data to send to a random destination, but in mesh networks, relays (mesh routers) do not generate data. 

Most research on the capacity or transmission rate of mesh networks, e.g., Jun \emph{et al.} \cite{junsichitiu03}\cite{jun02}, Kodialam and Nandagopal \cite{kodialam05}, Roy \emph{et al.}~\cite{roydas05}, is based on non-cooperative multihop coding strategies.
Data are forwarded hop-by-hop from the mesh clients to the gateway through relays and simultaneous transmissions use orthogonal channels, e.g., using time division multiple access (TDMA) or frequency division multiple access (FDMA), to avoid interference. In this paper, we give examples to show that the rate achievable by the TDMA multihop strategy is lower than that of the compress-and-forward strategy. The FDMA multihop strategy can be analyzed similarly.

The rest of the paper is organized as follows. Section~\ref{mesh_channel} introduces the channel model. In Section~\ref{mesh_theories}, we establish a useful theorem that we will need in later sections. In Section~\ref{mesh_upper_bound}, we investigate an upper bound on the capacity of the SSMRSD mesh network. This is followed by an investigation of the rates achievable by the compress-and-forward strategy, in Section~\ref{mehs_mrc}. By allowing the power constraint of the relays to increase, we show, in Section~\ref{mesh_capacity_ssmr}, that the rate achievable by the compress-and-forward strategy approaches the capacity asymptotically. In Section~\ref{sec:mesh_power_position}, we investigate the behavior of the compress-and-forward rate when the relay power is finite. In addition, we compare the performance of the compress-and-forward and the decode-and-forward strategies for different relay positions. We conclude this paper in Section~\ref{sec:mesh_conclusions}.

\section{Channel Model} \label{mesh_channel}
\begin{figure}[ht]
\centering
\includegraphics[width=6cm]{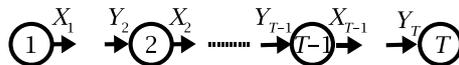}
\caption{The multiple relay channel.}  \label{fig:ssmr}
\end{figure}

Fig.~\ref{fig:ssmr} depicts the  multiple relay channel. The  multiple relay channel can be
completely described by the channel distribution
\begin{equation}
p^*(y_2, y_3, \dotsc, y_T | x_1, x_2, \dotsc, x_{T-1})
\end{equation}
on $\mathcal{Y}_2 \times \mathcal{Y}_3 \times \dotsm \times \mathcal{Y}_T$, for each $(x_1, x_2, \dotsc, x_{T-1}) \in \mathcal{X}_1 \times \mathcal{X}_2 \times \dotsm \times \mathcal{X}_{T-1}$.  In this paper, we only consider memoryless channels. Node 1 is the source node and node $T$ is the destination node.  Nodes 2 to $T-1$ are purely
relay nodes. Messages are generated at node 1 and are to be transferred to node $T$.
We follow the definitions of capacity, achievable rate ($R_W$) used by Kramer \emph{et al.} \cite[Section III.A]{kramergastpar04}.

In the Gaussian multiple relay channel, node $j$, $j=2, \dotsc, T$, receives
\begin{equation}
Y_j = \sum_{\substack{i=1, \dotsc, T-1 \\ i \neq j}} \sqrt{\lambda_{ij}}X_i + Z_j,
\end{equation}
where $X_i$, input to the channel from node $i$, is a random variable with power constraint $E[X_i^2] \leq P_i$. $P_i$ is the average transmission power constraint on node $i$. $Y_j$ is the received signal at node $j$. $Z_j$, the receiver noise at node $j$, is an independent zero mean Gaussian random variable with variance $N_j$. $\lambda_{ij}$ is the channel gain from node $i$ to node $j$. $\lambda_{ij}$ depends on the antenna gains, the carrier frequency of the transmission, and the distance between the transmitter and the receiver.

We study achievable rates for the $T$-node Gaussian multiple relay channel. More specifically, we investigate the performance of the compress-and-forward strategy when the relay powers grow large. This models the SSMRSD mesh network, in which the relays are line powered and can transmit at high power.
We define $\mathcal{R} \triangleq \{ 2, 3, \dotsc, T-1\}$ as the set of all relay nodes. We use the notation $X_{\{1, \dotsc, m\}} \triangleq (X_1, \dotsc, X_m)$.

\section{A Cut-Set Bound and Independent Gaussian Inputs} \label{mesh_theories}
In this section, we establish a useful theorem which we will need in the sequel. We consider a $T$-node multiple relay channel where nodes $1,\dotsc,T-1$ send $X_1, \dotsc, X_{T-1}$ into the channel respectively. The channel inputs are subject to power constraints $E[X_i] \leq P_i$ for $i=1,\dotsc,T-1$. Nodes $2, \dotsc, T$ receive the following signals from the channel.
\begin{equation}
Y_j = \sum_{i \in \{1\} \cup \mathcal{R} \setminus \{j\}} X_i + Z_j
\end{equation}
where $Z_j \sim \mathcal{N}(0,N_j)$, $j=2,3, \dotsc, T$, are independent Gaussian noise. Since the values of the channel gain do not matter in the analyses in this section, we have set them to be 1.

We consider the cut-set bound on the rate at which we can transmit information from the source to the relays and the destination, assuming that the relays and the destination can cooperate. The following theorem establishes that the optimal input distribution to maximize this bound is such that the the source and the relays send independent Gaussian inputs.

\begin{thm}\label{thm:mrc_gen_indep_gauss}
Consider a $T$-node Gaussian multiple relay channel. A sufficient condition on the input distribution that achieves
\begin{equation}\label{eq:mutual_mrc_gen_indep_gauss}
\max_{p(x_1,x_2, \dotsc, x_{T-1})} I(X_1;Y_\mathcal{R},Y_T|X_\mathcal{R})
\end{equation}
is that the inputs are Gaussian and $X_1$ is independent of $X_\mathcal{R}$. It follows that independent Gaussian inputs $X_1, \dotsc, X_{T-1}$ also achieve \eqref{eq:mutual_mrc_gen_indep_gauss}.
\end{thm}

\emph{Proof:}[Proof of Theorem~\ref{thm:mrc_gen_indep_gauss}] First, we consider the case $T=3$, which means there is one relay. We want to show that 
\begin{equation}
\max_{p(x_1,x_2)} I(X_1;Y_2,Y_3|X_2)
\end{equation}
is achieved when $X_1$ and $X_2$ are independent Gaussian inputs.

From \cite[Proposition 2]{kramergastpar04}, we know the optimal input distribution is Gaussian. We let
\begin{equation}
X_1 = \alpha X_2 + W,
\end{equation}
where $W$ and $X_2$ are independent Gaussian random variables, such that $E[W^2]=P_W$ and $P_1 = \alpha^2P_2 + P_W$.

Now,
\begin{equation}
H(Y_2,Y_3|X_1,X_2) = \frac{1}{2} \log (2\pi e)^2N_2N_3,
\end{equation}
and
\begin{subequations}
\begin{align}
H(Y_2,Y_3|X_2) & = \frac{1}{2} \log (2\pi e)^2 
\begin{vmatrix}
P_W + N_2 & P_W \\
P_W  & P_W + N_3
\end{vmatrix} \\
& = \frac{1}{2} \log (2\pi e)^2 (P_WN_2 + P_WN_3 + N_2N_3).
\end{align}
\end{subequations}

Hence,
\begin{subequations}
\begin{align}
I(X_1;Y_2,Y_3|X_2) & = H(Y_2,Y_3|,X_2) - H(Y_2,Y_3|X_1,X_2) \\
& = \frac{1}{2} \log \left[ 1 + \frac{P_1 - \alpha^2 P_2}{N_2} + \frac{P_1 - \alpha^2P_2}{N_3}  \right].
\end{align}
\end{subequations}
Setting $\alpha=0$ maximizes the mutual information. This completes the proof for $T=3$.

Now, we extend this result to $T=4$ or the two-relay channel. The generalization from the two-relay channel to the multiple-relay channel is straight forward. We need to show that a sufficient condition on the input distribution function to achieve
\begin{equation}\label{eq:mutual_mrc_indep_gauss}
\max_{p(x_1,x_2,x_3)} I(X_1;Y_2,Y_3,Y_4|X_2,X_3)
\end{equation}
is that $X_1$ and  $(X_2,X_3)$ are independent Gaussian inputs.

From \cite[Proposition 2]{kramergastpar04}, \eqref{eq:mutual_mrc_indep_gauss} is achieved by Gaussian inputs $X_1$, $X_2$, and $X_3$. Now, we combine the relay inputs to form $X_{\mathcal{R}} = X_2 + X_3$. The destination receives $Y_4 = X_1 + X_{\mathcal{R}} + N_4$. From the single relay case $T=3$, we know that choosing $X_1$ to be independent of $X_{\mathcal{R}}$ is optimal. Certainly, choosing independent $X_1$, $X_2$, and $X_3$ maximizes the mutual information term. This proves the case of $T=4$.

Now, we demonstrate that \eqref{eq:mutual_mrc_indep_gauss} can indeed be achieved with any correlation between $X_2$ and $X_3$, as long as $X_1$ is independent of $(X_2,X_3)$. We let $X_2 = \beta X_3 + W$, where $X_1$, $X_3$ and $W$ are independent Gaussian inputs. Here, $E[W^2]=P_W$ and $P_2 = \beta^2P_3 + P_W$.

Now,
\begin{equation}
H(Y_2,Y_3,Y_4|X_1,X_2,X_3) = \frac{1}{2} \log (2\pi e)^3 N_2N_3N_4.
\end{equation}
Also,
\begin{subequations}
\begin{align}
& H(Y_2,Y_3,Y_4|X_2,X_3) \nonumber \\
& = \frac{1}{2} \log (2\pi e)^3
\begin{vmatrix}
P_1+N_2 & P_1 & P_1 \\
P_1 & P_1+N_3 & P_1 \\
P_1 & P_1 & P_1+N_4
\end{vmatrix}\\
& = \frac{1}{2} \log (2\pi e)^3 \left[ P_1(N_2N_3+N_2N_4+N_3N_4) + N_2N_3N_4  \right].
\end{align}
\end{subequations}
Hence,
\begin{equation}
I(X_1;Y_2,Y_3,Y_4|X_2,X_3)  = \frac{1}{2} \log \left[ 1 + P_1\left( \frac{1}{N_2} + \frac{1}{N_3} + \frac{1}{N_4}  \right) \right].
\end{equation}
We note that this is independent of $\beta$. This means that \eqref{eq:mutual_mrc_indep_gauss} can be achieved with any correlation between $X_2$ and $X_3$.

We can easily generalize this result to any $T > 4$ and hence obtain Theorem~\ref{thm:mrc_gen_indep_gauss}.

\section{Capacity Upper Bounds} \label{mesh_upper_bound}
\subsection{The Multi-Terminal Network}
\begin{figure}[ht]
\centering
\includegraphics[width=4cm]{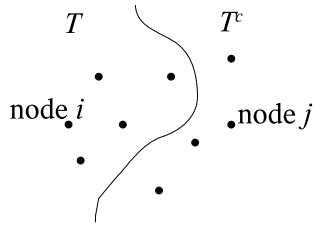}
\caption{A cut in the multi-terminal network.}  \label{fig:cut_multi_terminal}
\end{figure}

Consider a $T$-node multi-terminal network where node $i$ transmits $X_i$ and node $j$ receives $Y_j$. The channel is characterized by the channel transition probability $p(y_1, \dotsc, y_T | x_1, \dotsc, x_T)$.  If the rate from node $i$ to node $j$, $R_{ij}$, is achievable, then the following must be satisfied \cite[Theorem 14.10.1]{coverthomas91}
\begin{equation}\label{eq:cut_rate}
\sum_{i \in \mathcal{T}, j \in \mathcal{T}^c} R_{ij} \leq \max_{p(x_1, \dotsc, x_T)}I(X_\mathcal{T} ; Y_{\mathcal{T}^c} | X_{\mathcal{T}^c} ),
\end{equation}
for some joint probability function $p(x_1, \dotsc, x_T)$ for all $\mathcal{T} \subset \{ 1, \dotsc, T\}$ where $i \in \mathcal{T}$ and $j \notin \mathcal{T}$. $\mathcal{T}^c$ is the complement of $\mathcal{T}$ in $\{1, \dotsc, T\}$.

We can interpret this theorem as follows. The achievable rate from node $i$ to node $j$ must be smaller than the rate of all possible cuts separating nodes $i$ and $j$. Fig.~\ref{fig:cut_multi_terminal} depicts a possible cut. We define the \emph{cut rate} for the cut separating $\mathcal{T}$ and $\mathcal{T}^c$ as the right side of \eqref{eq:cut_rate}. It is the maximum achievable rate from nodes in $\mathcal{T}$ to nodes in $\mathcal{T}^c$ when all nodes on the same side of the cut are allowed to cooperate.

\subsection{The SSMRSD Mesh Network}
Consider a $T$-node Gaussian SSMRSD mesh network where
\begin{itemize}
	\item Node 1 is the source node with power constraint $E[X_1^2] \leq P_1$. The source only transmits and does not receive.
	\item Node $T$ is the destination node, which only receives and does not transmit.
	\item Nodes 2 to $T-1$ are relays with large power constraints, $E[X_i^2] \leq P_i$ and $P_i >> P_1$, $\forall i = 2, \dotsc, T-1$. The relays can transmit and receive at the same time.
\end{itemize}
We note that any cut rate with $1 \in \mathcal{T}$ and $T \in \mathcal{T}^c$ is an upper bound of the rate from the source to the destination. Since the relays can transmit with large power, if we include any relay node in set $\mathcal{T}$, the cut rate (defined as \eqref{eq:cut_rate}) is large. Hence the minimum cut rate occurs when the cut separates $\mathcal{T} = \{ 1 \}$ and $\mathcal{T}^c = \{2, \dotsc, T \}$. So the upper bound of the capacity of the SSMRSD reduces to
\begin{equation}\label{eq:cutset_mesh}
C_{\text{SSMRSDMesh}} \leq \max_{\substack{p(x_1, \dotsc, x_{T-1})}} I(X_1 ; Y_\mathcal{R},Y_T | X_\mathcal{R}),
\end{equation}
for some joint probability function $p(x_1, \dotsc, x_{T-1})$. From Theorem~\ref{thm:mrc_gen_indep_gauss}, independent Gaussian inputs maximize this upper bound in the Gaussian channel.

\section{Achievable Rates} \label{mehs_mrc}

\subsection{The Discrete Memoryless Multiple Relay Channel}
In this section, we investigate achievable rates of the multiple relay channel using the compress-and-forward strategy. Setting $U_t = X_t$, $\forall t \in \mathcal{R}$ in the compress-and-forward strategy for the multiple relay channel introduced by Kramer \emph{et al.} \cite[Theorem 3]{kramergastpar04}, we can achieve rates up to

\begin{equation}\label{eq:cf_rate_mesh}
R = I(X_1; \tilde{Y}_\mathcal{R}, Y_T | X_\mathcal{R}),
\end{equation}
where
\begin{equation}\label{eq:cf_condition_mesh}
I(\tilde{Y}_\mathcal{S} ; Y_\mathcal{S} | X_\mathcal{R}, \tilde{Y}_{\mathcal{S}^c}, Y_T) \leq \sum_{m=1}^M I( X_{\mathcal{B}_m} ; Y_{r(m)} | X_{\mathcal{B}_m^c}),
\end{equation}
with the joint probability distribution function
\begin{equation}\label{eq:cf_distribution_mesh}
p(x_1) \left[ \prod_{t \in \mathcal{R}}p(x_t)p(\tilde{y}_t | x_{\mathcal{R}},y_t) \right] p^*(y_\mathcal{R}, y_T | x_1, x_\mathcal{R}).
\end{equation}

for all $\mathcal{S} \subseteq \mathcal{R}$, all partitions $\{\mathcal{B}_m \}_{m=1}^M$ of $\mathcal{S}$, and all $r(m) \in \{ 2, \dotsc, T\} \setminus \mathcal{B}_m$. We define $\mathcal{S}^c$ as the complement of $\mathcal{S}$ in $\mathcal{R}$ and $\mathcal{B}_m^c$ as the compliment of $\mathcal{B}_m$ in $\mathcal{R}$. $U$ is the part which is to be decoded by all relays. Setting $U_t = X_t$ means each relay decodes all other relays' codewords. We note that in the compress-and-forward strategy, all channel inputs $X_1, \dotsc, X_{T-1}$ are independent.

\subsection{The Gaussian Multiple Relay Channel}

We consider the Gaussian multiple relay channel. The inputs to the channel from node $i$, $X_i$, is constrained by $E[X_i] \leq P_i$. We let the input distribution to the channel be independent Gaussian. Using the compress-and-forward strategy with $U_j = X_j$, the received signal of node $r(m)$ can be written as
\begin{subequations}
\begin{align}
Y_{r(m)} & = \sqrt{\lambda_{1r(m)}}X_1 + \sum_{\substack{i \in \mathcal{R} \\ i \neq r(m)}} \sqrt{\lambda_{ir(m)}}X_i + Z_{r(m)}\\
& = \sqrt{\lambda_{1r(m)}}X_1 + \sum_{\substack{i \in \mathcal{B}_m \\ i \neq r(m)}} \sqrt{\lambda_{ir(m)}}X_i + \sum_{\substack{i \in \mathcal{B}_m^c \\ i \neq r(m)}} \sqrt{\lambda_{ir(m)}}X_i \nonumber\\
& \quad + Z_{r(m)}.
\end{align}
\end{subequations}

The term inside the summation on the RHS of \eqref{eq:cf_condition_mesh} can be evaluated as
\begin{equation}
I( X_{\mathcal{B}_m} ; Y_{r(m)} | X_{\mathcal{B}_m^c}) = \frac{1}{2} \log \left[ 1 + \frac{\sum_{\substack{i \in \mathcal{B}_m \\ i \neq r(m)}} \lambda_{ir(m)}P_i}{\lambda_{1r(m)}P_1 + N_{r(m)}} \right].
\end{equation}
We note that all $X_i$ are independent, as seen from \eqref{eq:cf_distribution_mesh}.

Using the compress-and-forward strategy, node $j$'s quantized received signal is
\begin{equation}
\tilde{Y}_j = Y_j + W_j = \sum_{\substack{i=1, \dotsc, T-1 \\ i \neq j}} \sqrt{\lambda_{ij}}X_i + Z_j + W_j,
\end{equation}
where $W_j \sim \mathcal{N}(0,Q_j)$ are independent quantization noise.

The LHS of \eqref{eq:cf_condition_mesh} is
\begin{subequations}
\begin{align}
I(\tilde{Y}_\mathcal{S} ; Y_\mathcal{S} | X_\mathcal{R}, \tilde{Y}_{\mathcal{S}^c}, Y_T)
& = H(\tilde{Y}_\mathcal{S} | X_\mathcal{R}, \tilde{Y}_{\mathcal{S}^c}, Y_T) - H(\tilde{Y}_\mathcal{S} | Y_\mathcal{S}, X_\mathcal{R}, \tilde{Y}_{\mathcal{S}^c}, Y_T) \\
& = H(\tilde{Y}_\mathcal{S} | X_\mathcal{R}, \tilde{Y}_{\mathcal{S}^c}, Y_T) - H(W_\mathcal{S})\\
& \leq H(\tilde{Y}_\mathcal{S} | X_\mathcal{R}) - H(W_\mathcal{S}). \label{eq:bound}
\end{align}
\end{subequations}
The first term in \eqref{eq:bound} is
\begin{equation}
H(\tilde{Y}_\mathcal{S} | X_\mathcal{R}) = \frac{1}{2} \log \left[ 2\pi e^D \Lambda(D) \right],
\end{equation}
where $\Lambda(D)$ is defined as follows,

\begin{equation}\label{eq:determinant}
\Lambda(D) =
\begin{vmatrix}
\lambda_{1s(1)}P_1 + N_{s(1)} + Q_{s(1)} & \dotsm & \sqrt{\lambda_{1s(1)}\lambda_{1s(D)}}P_1 \\
\vdots & \ddots & \vdots \\
\sqrt{\lambda_{1s(1)}\lambda_{1s(D)}}P_1 & \dotsc & \lambda_{1s(D)}P_1 + N_{s(D)} + Q_{s(D)}
\end{vmatrix}
\end{equation}
$s(i)$ are the ordered elements in $\mathcal{S}$ and $D = \lvert \mathcal{S} \rvert$.

The second term in \eqref{eq:bound} is
\begin{equation}
H(W_\mathcal{S}) = \frac{1}{2} \log \left[ 2\pi e^D Q_{s(1)} \dotsm Q_{s(D)} \right].
\end{equation}

Now a sufficient condition for \eqref{eq:cf_condition_mesh} is 
\begin{equation}
H(\tilde{Y}_\mathcal{S} | X_\mathcal{R}) - H(W_\mathcal{S}) \leq \sum_{m=1}^M I( X_{\mathcal{B}_m} ; Y_{r(m)} | X_{\mathcal{B}_m^c}),
\end{equation}
or in the Gaussian channel,
\begin{equation}
Q_{s(1)} \dotsm Q_{s(D)} \geq \frac{\Lambda(D)}{\prod_{m=1}^M \left[ 1 + \frac{\sum_{\substack{i \in \mathcal{B}_m \\ i \neq r(m)}} \lambda_{ir(m)}P_i}{\lambda_{1r(m)}P_1 + N_{r(m)}} \right]}.
\end{equation}

The achievable rate is given by
\begin{subequations}
\begin{align}
I(X_1; \tilde{Y}_\mathcal{R}, Y_T | X_\mathcal{R}) & = H(\tilde{Y}_\mathcal{R}, Y_T | X_\mathcal{R}) - H(\tilde{Y}_\mathcal{R}, Y_T | X_1, X_\mathcal{R}) \\
& = \frac{1}{2} \log \left[ 2\pi e^{T-1} \Psi(T-1) \right] \nonumber\\
& \quad - \frac{1}{2} \log \left[ 2\pi e^{T-1}(N_2+Q_2) \dotsm (N_{T-1} + Q_{T-1})N_T \right] \\
& = \frac{1}{2} \log \left[ 1 + \frac{\lambda_{12} P_1}{N_2+Q_2} + \dotsm + \frac{\lambda_{1T-1} P_1}{N_{T-1}+Q_{T-1}} + \frac{\lambda_{1T} P_1}{N_T}  \right],
\end{align}
\end{subequations}
where
\begin{equation}
\Psi(T-1) =
\begin{vmatrix}
\lambda_{12}P_1 + N_{2} + Q_{2} & \dotsm & \sqrt{\lambda_{12}\lambda_{1T}}P_1 \\
\vdots & \ddots & \vdots \\
\sqrt{\lambda_{12}\lambda_{1T}}P_1 & \dotsc & \lambda_{1T}P_1 + N_{T}
\end{vmatrix}.
\end{equation}

Hence we have the following theorem for the $T$-node Gaussian multiple relay channel.

\begin{thm}\label{thm:cf_mesh_achievable}
Consider a memoryless $T$-node Gaussian multiple relay channel. Using independent Gaussian input $X_i$, $i=1, \dotsc, T-1$, with power constraints $E[X_i^2] \leq P_i$, the compress-and-forward strategy achieves rates up to
\begin{equation}\label{eq:rate_achievable_mesh_thm}
R = \frac{1}{2} \log \left[ 1 + \frac{\lambda_{12} P_1}{N_2+Q_2} + \dotsm + \frac{\lambda_{1T-1} P_1}{N_{T-1}+Q_{T-1}} + \frac{\lambda_{1T} P_1}{N_T}  \right],
\end{equation}
The rate equation is subject to the constraints
\begin{equation}\label{eq:ssmrmesh_cap_condition}
Q_{s(1)} \dotsm Q_{s(D)} \geq \frac{\Lambda(D)}{\prod_{m=1}^M \left[ 1 + \frac{\sum_{\substack{i \in \mathcal{B}_m \\ i \neq r(m)}} \lambda_{ir(m)}P_i}{\lambda_{1r(m)}P_1 + N_{r(m)}} \right]},
\end{equation}
for all $\mathcal{S} \subseteq \mathcal{R}$, $\{s(1)... s(D)\}=\mathcal{S}$, all partitions $\{\mathcal{B}_m \}_{m=1}^M$ of $\mathcal{S}$, and all $r(m) \in \{ 2, \dotsc, T\} \setminus \mathcal{B}_m$. $\mathcal{R}$ is the set of all relays.
\end{thm}

We note that the achievability of \eqref{eq:rate_achievable_mesh_thm} makes use of the Markov lemma~\cite[Lemma 4.1]{berger77}, which requires strong typicality. Though strong typicality does not extend to continuous random variables, one can generalize the Markov lemma for Gaussian inputs and thus show that \eqref{eq:rate_achievable_mesh_thm} is achievable \cite{kramergastpar04}.

\section{The Capacity of the Gaussian SSMRSD Mesh Network} \label{mesh_capacity_ssmr}
Recalling that the Gaussian SSMRSD mesh network is a multiple relay channel with powerful relays, we investigate how \eqref{eq:ssmrmesh_cap_condition} can be satisfied for some small $Q_i$ in the mesh network.
To see the effect of these powerful nodes, we will look at the scenario when the relay power constraints grow without bound, i.e., $P_i \rightarrow \infty, \forall i \in \mathcal{R}$.
While this may not be practical, it allows us to characterize the capacity and to study how the rates scale with the relay powers.
We assume that all the channel gains $\lambda_{ij}$, the transmit power $P_1$, and the receiver noise $N_i$ are finite.
Under these conditions:
\begin{enumerate}
\item $\Lambda(D)$ is finite when $P_1$, $N_i$, $\lambda_{1i}$ are finite and $Q_i$ are approaching zero, $\forall i \in \mathcal{R}$.
\item $(\lambda_{1j}P_1 + N_{j}), \forall j \in \mathcal{R} \cup \{ T \}$ are finite.
\item The RHS of \eqref{eq:ssmrmesh_cap_condition} approaches zero as $\lambda_{ij}P_i \rightarrow \infty, \forall i \in \mathcal{R}, \forall j \in \mathcal{R} \cup \{ T \}$.
\end{enumerate}

Hence, we can set
\begin{equation}
Q_i \rightarrow 0, \forall i \in \mathcal{R},
\end{equation}
while \eqref{eq:ssmrmesh_cap_condition} can still be satisfied  for all $\mathcal{S} \subseteq \mathcal{R}$, all partitions $\{\mathcal{B}_m \}_{m=1}^M$ of $\mathcal{S}$, and all $r(m) \in \{ 2, \dotsc, T\} \setminus \mathcal{B}_m$.
When $Q_i \rightarrow 0$, the quantized received signals approach the received signals, that is
\begin{equation}
\tilde{Y}_i = Y_i + W_i \rightarrow Y_i,
\end{equation}
for all $\forall i \in \mathcal{R}$. 
The achievable rate of the compress-and-forward strategy becomes
\begin{equation}\label{eq:rate_achievable_mesh}
R \rightarrow \max_{\substack{\text{independent Gaussian inputs}\\E[X_1^2] \leq P_1}} I(X_1; Y_\mathcal{R}, Y_T | X_\mathcal{R}).
\end{equation}

We see that \eqref{eq:rate_achievable_mesh} has the same form as the capacity upper bound \eqref{eq:cutset_mesh} of the SSMRSD mesh network. The upper bound \eqref{eq:cutset_mesh} is maximized is over all possible input distributions but the achievable rate \eqref{eq:rate_achievable_mesh} is achievable with independent Gaussian inputs. However, Theorem~\ref{thm:mrc_gen_indep_gauss} states that the cut-set upper bound is maximized by using independent Gaussian inputs. Hence, the compress-and-forward strategy approaches a cut-set upper bound of the SSMRSD mesh network asymptotically.  This is summarized in the following theorem.
\begin{thm} \label{thm:mesh_cap}
The achievable rate of the compress-and-forward strategy approaches the capacity of the Gaussian SSMRSD mesh network asymptotically as the relay powers grow. The capacity is given by
\begin{equation}\label{eq:ssmr_mesh_cap_thm}
C_{\text{SSMRSDMesh}} = \max_{\substack{\text{independent Gaussian inputs}\\E[X_1^2] \leq P_1}}I(X_1; Y_\mathcal{R}, Y_T | X_\mathcal{R}).
\end{equation}
\end{thm}

Similar to Theorem~\ref{thm:cf_mesh_achievable}, the achievability of \eqref{eq:ssmr_mesh_cap_thm} requires generalization of the Markov lemma to Gaussian input distributions \cite{kramergastpar04}.

\section{The Effect of Relay Position and Power} \label{sec:mesh_power_position}

%\subsection{Preliminary}

In the previous section, we have shown that in the multiple relay channel when the relay powers get large, the achievable rate of the compress-and-forward strategy approaches the capacity. While the assumption that the relay powers grow to infinity is not valid in any practical scenario, we investigate the case of finite relay powers in this section. We study, using the single relay channel as an example, how close the achievable rate of the compress-and-forward strategy gets to the capacity when we increase the relay power. In addition, we investigate how the relay position affects the result.

\subsection{Comparing Different Coding Strategies and the Cut-Set Bound}
We recall that the cut-set bound for the single relay channel is
\begin{subequations}
\begin{align}
R_{\text{CS}} & = \max_{p(x_1,x_2)} \min \{ I(X_1; Y_2, Y_3 | X_2), I(X_1,X_2; Y_3) \} \\
& = \max_{0 \leq \alpha_{\text{CS}} \leq 1} \min \left\{ L\left( \left(\frac{P_1\lambda_{13}}{N_3} + \frac{P_1 \lambda_{12}}{N_2} \right) (1 - \alpha_{\text{CS}}) \right), \right. \nonumber \\
& \quad \quad \left. L\left( \frac{P_1\lambda_{13}}{N_3} + \frac{P_2\lambda_{23}}{N_3} + \frac{2 \sqrt{\alpha_{\text{CS}}  \lambda_{13} \lambda_{23} P_1 P_2 }}{N_3}    \right) \right\},
\end{align}
\end{subequations}
where $L(x) = \frac{1}{2} \log (1 + x)$. In the numerical analyses in this section, we use this cut-set bound as a basis of comparison. We compare achievable rates of different coding strategies normalized to this bound.

The maximum achievable rate of the decode-and-forward strategy is~\cite{covergamal79}
\begin{subequations}
\begin{align}
R_{\text{DF}} & = \max_{p(x_1,x_2)} \min \{ I(X_1;Y_2 | X_2) , I(X_1, X_2; Y_3) \} \\
& = \max_{0 \leq \alpha_{\text{DF}} \leq 1} \min \left\{ L\left( \frac{P_1 \lambda_{12}}{N_2}(1 - \alpha_{\text{DF}}) \right), \right. \nonumber \\
& \quad \quad \left. L\left( \frac{P_1\lambda_{13}}{N_3} + \frac{P_2\lambda_{23}}{N_3} + \frac{2 \sqrt{\alpha_{\text{DF}}  \lambda_{13} \lambda_{23} P_1 P_2 }}{N_3}    \right) \right\}.
\end{align}
\end{subequations}
In the decode-and-forward strategy, the source transmits to both the relay and the destination. The relay fully decodes the data sent by the source, and helps the source to transmit to the destination.

The maximum achievable rate of the compress-and-forward strategy is
\begin{subequations}
\begin{align}
R_{\text{CF}} & = I(X_1;\tilde{Y}_2, Y_3 | X_2), \quad \text{where } I(X_2;Y_3) \geq I(Y_2; \tilde{Y}_2|X_2,Y_3) \\
& = L\left( \frac{P_1 \lambda_{13}}{N_3} + \frac{P_2 \lambda_{12}}{N_2 + Q} \right), \\& \quad \quad \text{where } Q = \frac{(\lambda_{13}N_2 + \lambda_{12}N_3)P_1 + N_2N_3}{P_2 \lambda_{23}}.
\end{align}
\end{subequations}

The maximum achievable rate of the TDMA multihop strategy is
\begin{equation}
R_{MH} = \max_{0 \leq \alpha_{MH} \leq 1} \min \left\{  (1-\alpha_{MH}) L \left( \frac{P_1\lambda_{12}}{(1-\alpha_{MH})N_2} \right), \alpha_{MH} L \left( \frac{P_2\lambda_{23}}{\alpha_{MH}N_3} \right) \right\}.
\end{equation}
In this strategy, during $(1-\alpha_{MH})$ fraction of the time, the source transmits to the relay; during $\alpha_{MH}$ fraction of the time, the relay transmits to the destination. This is the strategy proposed in \cite{junsichitiu03}\cite{jun02}.

\subsection{Channel Models}
In order to compare different strategies at different relay positions and powers, we need to select a model for the channel gain $\lambda_{ij}$ between two nodes $i$ and $j$. Using the Friis free space path loss model, the channel gain in an unobstructed line-of-sight environment is given by
\begin{equation}
\lambda_{ij} = \frac{P_j}{P_i} = \frac{G}{(4 \pi f d_{ij})^2},
\end{equation}
where $P_j$ is the received power, $P_i$ the transmit power, $G$ the antenna gain, $f$ is the carrier frequency, and $d_{ij}$ the distance between the transmitter and the receiver. In other environments, there are different models for signal propagation attenuation. However, in most models, the channel gain is proportional to $d_{ij}^{-\eta}$, where $\eta$ ranges from 2 to 8. Capturing the main characteristic of how the channel gain varies with distance, one can simplify these models to the following simplified path loss (SPL) model.
\begin{equation}
\lambda_{ij} = \kappa d_{ij}^{-\eta},
\end{equation}
where $\eta$ is the path loss exponent, and $\eta \geq 2$ with equality for free space transmission. $\kappa$ is a positive constant as far as the analyses in this section are concerned. The SPL model is a widely accepted model and commonly used in the information theoretic literature ~\cite{kramergastpar04}\cite{guptakumar00}\cite{toumpisgoldsmith03}\cite{gatsparvetterli05}.

However, the SPL model is only valid for distances in the far field. At small distances, the SPL model diverges, i.e., as $d_{ij} \rightarrow 0$, $\lambda_{ij} \rightarrow \infty$. Near field signal propagation models~\cite{ravindrasarma99} suffer the same problem when we consider asymptotic cases when the distance approaches zero. We now consider how to modify the SPL model to rectify this problem.
We note that a reasonable signal propagation model should have the following properties.
\begin{enumerate}
\item The channel gain is bounded $\forall d_{ij} \geq 0$.
\item The channel gain decreases monotonically as $d_{ij}$ increases.
\item The channel gain $\lambda_{ij} \rightarrow \kappa d_{ij}^{-\eta}$ as $d_{ij} \rightarrow \infty$.
\end{enumerate}

Keeping these in mind, we propose the modified path loss (MPL) model.
\begin{equation}
\lambda_{ij} = \kappa (1 + d_{ij})^{-\eta}.
\end{equation}
Though the MPL model might not be accurate for general environments, it captures the above mentioned properties of signal propagation. It fixes the undesirable behavior of the SPL model when $d_{ij} \rightarrow 0$, and allows us to compare different coding strategies with different relay positions, especially in this extreme condition.

\begin{remark}
For the single relay channel~\cite{kramergastpar04} and the multiple relay channel~\cite{gastparkramer02}, it has been noted that when the relay(s) moves toward the destination, the compress-and-forward strategy approaches the capacity. In all these scenarios the relay powers are finite.
At a first look, that the relay is close to the destination seems equivalent to the case when the relay power gets large. In this section, we will show that they are not equivalent. We note that the compress-and-forward strategy achieves the capacity when the relay approaches the destination because the channel gain from the relay to the destination approaches infinity using the SPL model for signal propagation. As discussed earlier, the SPL model is not valid when the inter-node distance approaches zero. We show that the compress-and-forward strategy does not approach the capacity as the relay-destination distance goes to zero using another model, e.g., the MPL model, in which the channel gain is bounded. However, Theorem~\ref{thm:mesh_cap} holds regardless of the choice of path loss model.
\end{remark}

\subsection{Scenarios for Comparison}

Now, we investigate how the two factors, i.e., the relay power and the relay position, affect the achievable rates of the coding strategies. We consider the single relay channel for illustration. We assume that the source, the relay, and the destination are located in a straight line with the relay placed between the source and the destination. We fix the source-destination distance $d_{13}=1$. Now, let us consider eight cases, using different path loss models, for different relay powers $P_2$, relay-source distances $d_{12}$, and relay-destination distances $d_{23}$.
\begin{enumerate}
\item  SPL: $P_2 \rightarrow \infty, d_{12} \rightarrow 0, \lambda_{12} = \kappa d_{12}^{-\eta} \rightarrow \infty$,
\begin{enumerate}
\item \label{lab:case4} $\frac{P_2}{\lambda_{12}} \rightarrow \infty$,
\item \label{lab:case5} $\frac{P_2}{\lambda_{12}} = K_3$,
\item \label{lab:case6} $\frac{P_2}{\lambda_{12}} \rightarrow 0$,
\end{enumerate}
\item \label{lab:case7} SPL: $P_2 \rightarrow \infty, d_{12} = K_2$,
\item \label{lab:case8} SPL: $P_2 \rightarrow \infty, d_{23} \rightarrow 0$,
\item \label{lab:case1} SPL: $P_2 = K_1, d_{12} \rightarrow 0$,
\item \label{lab:case2} SPL: $P_2 = K_1, d_{12} = K_2$,
\item \label{lab:case3} SPL: $P_2 = K_1, d_{23} \rightarrow 0$,
\item \label{lab:case9} MPL: $P_2 \rightarrow \infty$
\item \label{lab:case10} MPL: $P_2 = K_1$
\end{enumerate}
for some $K_1 \approx P_1$, $K_2 \approx 0.5$, $K_3 \approx 1$.

To understand the performance of the different strategies with varying parameters, we compute the gap between the achievable rates of the strategies and the cut-set bound.  These results are shown in Figs. \ref{fig:relay_close_to_destination_vary_distance}-\ref{fig:relay_close_to_destination_vary_power_mpl} and discussed below.  Note that we normalize the values and take logarithms for ease of comparison and visualization.

\subsection{Computations with the SPL Model}

\begin{table}[ht]
\caption{The effect of the relay distance and power on achievable rates and the capacity $\mathcal{C}$, in the SPL model.}
\label{tab:relay_distance_power}
\centering
\scriptsize
\begin{tabular}{| c  || c | c | c | c | c |}
\hline
 &  \multicolumn{3}{c |}{$d_{12} \rightarrow 0$} & $d_{12} \approx 0.5$ & $d_{23} \rightarrow 0$ \\
\cline{2-4}
 & $\frac{P_2}{d_{12}^{-\eta}} \rightarrow \infty$  & $\frac{P_2}{d_{12}^{-\eta}} = K_3$ & $\frac{P_2}{d_{12}^{-\eta}} \rightarrow 0$ & & \\
\hline \hline
$P_2 \rightarrow \infty$
& $\begin{array}{cccc}
 \text{case \eqref{lab:case4}} \\ R_{\text{DF}} \rightarrow \mathcal{C} (d_{12} \downarrow) \\ R_{\text{DF}} \approx \mathcal{C} (P_2 \uparrow)\\ R_{\text{CF}} \rightarrow \mathcal{C}
\end{array}$
& $\begin{array}{cccc}
 \text{case \eqref{lab:case5}} \\ R_{\text{DF}} \rightarrow \mathcal{C} (d_{12} \downarrow) \\ R_{\text{DF}} \nrightarrow \mathcal{C} (P_2 \uparrow)\\ R_{\text{CF}} \quad \mbox{Unknown}
\end{array}$
& $\begin{array}{ccccc}
 \text{case \eqref{lab:case6}} \\ R_{\text{DF}} \rightarrow \mathcal{C} (d_{12} \downarrow) \\ R_{\text{DF}} \nrightarrow \mathcal{C} (P_2 \uparrow)\\ R_{\text{CF}} \nrightarrow \mathcal{C} (d_{12} \downarrow) \\
R_{\text{CF}} \rightarrow \mathcal{C} (P_2 \uparrow)
\end{array}$
& $\begin{array}{ccc}
 \text{case \eqref{lab:case7}} \\ R_{\text{DF}} < \mathcal{C} \\ R_{\text{CF}} \rightarrow \mathcal{C}
\end{array}$
& $\begin{array}{ccc}
 \text{case \eqref{lab:case8}} \\ R_{\text{DF}} < \mathcal{C} \\ R_{\text{CF}} \rightarrow \mathcal{C}
\end{array}$ \\
\hline
$P_2 < \infty$
& \multicolumn{3}{c |}{ $\begin{array}{ccc}
 \text{case \eqref{lab:case1}} \\ R_{\text{DF}} \rightarrow \mathcal{C} \\ R_{\text{CF}} < \mathcal{C}
\end{array}$
 } & $\begin{array}{ccc}
 \text{case \eqref{lab:case2}} \\ R_{\text{DF}} < R_{\text{CS}} \\ R_{\text{CF}} < R_{\text{CS}}
\end{array}$ & $\begin{array}{ccc}
 \text{case \eqref{lab:case3}} \\ R_{\text{DF}} < \mathcal{C} \\ R_{\text{CF}} \rightarrow \mathcal{C}
\end{array}$\\
\hline
\end{tabular}
\vspace{1cm}
\end{table}

\begin{figure}[t]
\begin{minipage}[t]{0.48\linewidth}
\centering
\includegraphics[width=1\textwidth]{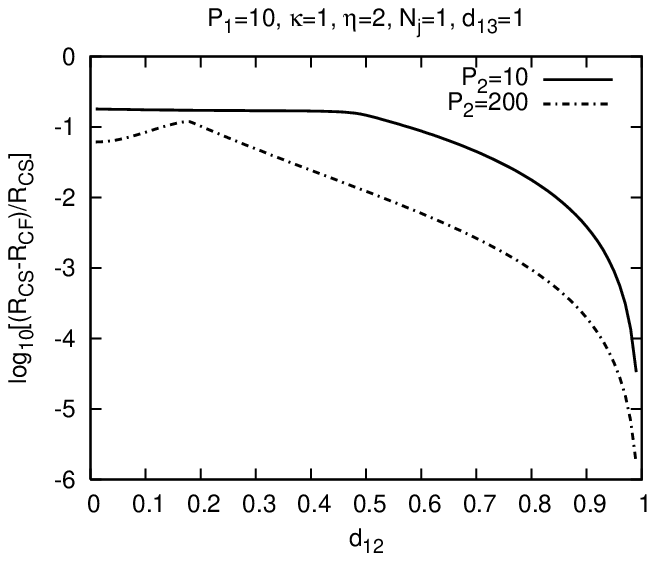}
\caption{The performance of the compress-and-forward strategy in the SPL model.}
\label{fig:relay_close_to_destination_vary_distance}
\end{minipage}
\hfill
\begin{minipage}[t]{0.48\linewidth}
\centering
\includegraphics[width=1\textwidth]{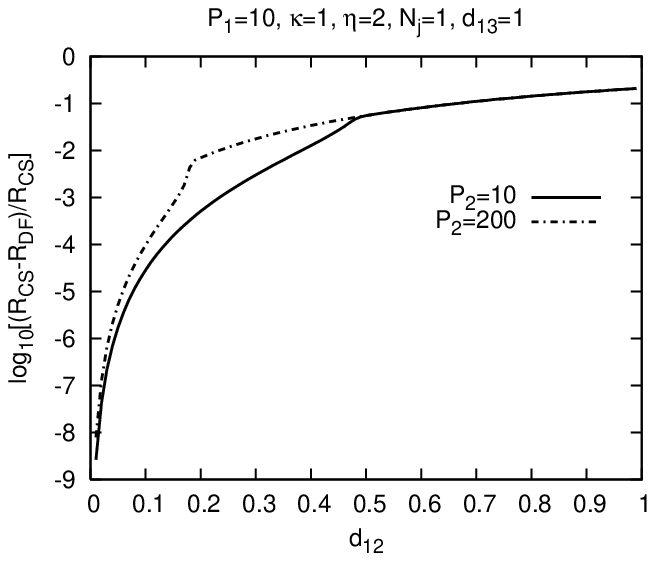}
\caption{The performance of the decode-and-forward strategy in the SPL model.}
\label{fig:relay_close_to_source_vary_distance_df}
\end{minipage}
\end{figure}

Table~\ref{tab:relay_distance_power} summarizes the results for cases~\eqref{lab:case4}--\eqref{lab:case3}, i.e., those using the SPL model. The proof for these cases can be found in Appendix~\ref{append:relay_distance_power}. The cases when the relay power is not large has been considered~\cite{kramergastpar04} but we include them here for completeness sake. 
As shown in Figs. \ref{fig:relay_close_to_destination_vary_distance} and \ref{fig:relay_close_to_source_vary_distance_df}, the decode-and-forward strategy achieves the capacity when the relay-source distance goes to zero and the compress-and-forward strategy approaches the capacity when the relay-destination distance goes to zero. These happen because $\lambda_{12} \rightarrow \infty$ as $d_{12} \rightarrow 0$, and $\lambda_{23} \rightarrow \infty$ as $d_{23} \rightarrow 0$ in the SPL model. In the MPL model, we will see that these strategies do not approach capacity with low $P_2$ regardless of the relay position.

%However, when the relay's power is large, the situation gets complicated. The compress-and-forward strategy still achieves the capacity asymptotically when the relay is near the destination. It also approaches the capacity for relay not near both the source and the destination. When the relay moves nearer to the source, with high relay's power, different strategies approach the capacity depending on the $\frac{P_2}{\lambda_{12}}$ ratio. Furthermore, different strategies approach the capacity, and possibly at a different convergence rate, when we vary $P_2$ or $d_{12}$.

%Figs. \ref{fig:relay_close_to_destination_vary_distance} and \ref{fig:relay_close_to_source_vary_distance_df} show the performance of the compress-and-forward strategy when the relay is close to the destination and the performance of the decode-and-forward strategy when the relay is close to the source, respectively. We see that as the relay-destination distance $d_{23}$ diminishes, or equivalently, $d_{12} \rightarrow 1$, the rate achievable by the compress-and-forward strategy approaches the capacity. This corresponds to case~\eqref{lab:case3}. When the source-the relay's distance $d_{12}$ diminishes, or equivalently, $d_{23} \rightarrow 1$, the rate achievable by the decode-and-forward strategy approaches the capacity. This corresponds to case~\eqref{lab:case1}.

%Figs.~\ref{fig:relay_close_to_destination_vary_power} and \ref{fig:relay_close_to_source_vary_power_cf} show that the compress-and-forward strategy approaches the capacity regardless of the relay's position as long as $P_2$ grows large. These correspond to cases~\eqref{lab:case4}, \eqref{lab:case7}, and \eqref{lab:case8}.

\subsection{Computations with the MPL Model}

\begin{table}[t]
\caption{The effect of the relay distance and power on achievable rates and the capacity $\mathcal{C}$ in the MPL model.}
\label{tab:relay_distance_power_mpl}
\centering
\scriptsize
\begin{tabular}{| c  || c | c | c |}
\hline
 &  $d_{12} \rightarrow 0$ & $d_{12} \approx 0.5$ & $d_{23} \rightarrow 0$ \\
\hline \hline
$P_2 \rightarrow \infty$
& \multicolumn{3}{c |}{ $\begin{array}{c}
\text{case \eqref{lab:case9}} \\ R_{\text{DF}} < \mathcal{C} \\ 
R_{\text{CF}} \rightarrow \mathcal{C} (P_2 \uparrow)
\end{array}$}\\
\hline
$P_2 < \infty$
& \multicolumn{3}{c |}{ $\begin{array}{c}
\text{case \eqref{lab:case10}} \\ R_{\text{DF}} < R_{\text{CS}} \\ 
R_{\text{CF}} < R_{\text{CS}}
\end{array}$}\\
\hline
\end{tabular}
\vspace{1cm}
\end{table}

\begin{figure}[t]
\begin{minipage}[t]{0.48\linewidth}
\centering
\includegraphics[width=1\textwidth]{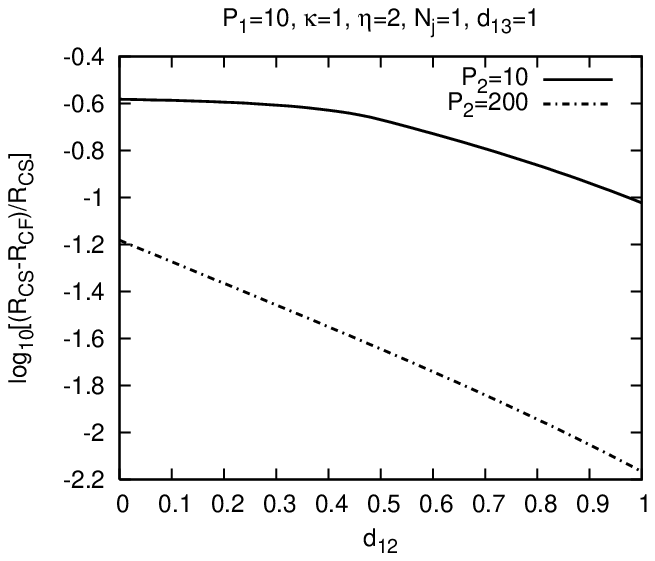}
\caption{The performance of the compress-and-forward strategy in the MPL model.}
\label{fig:relay_close_to_destination_vary_distance_mpl}
\end{minipage}
\hfill
\begin{minipage}[t]{0.48\linewidth}
\centering
\includegraphics[width=1\textwidth]{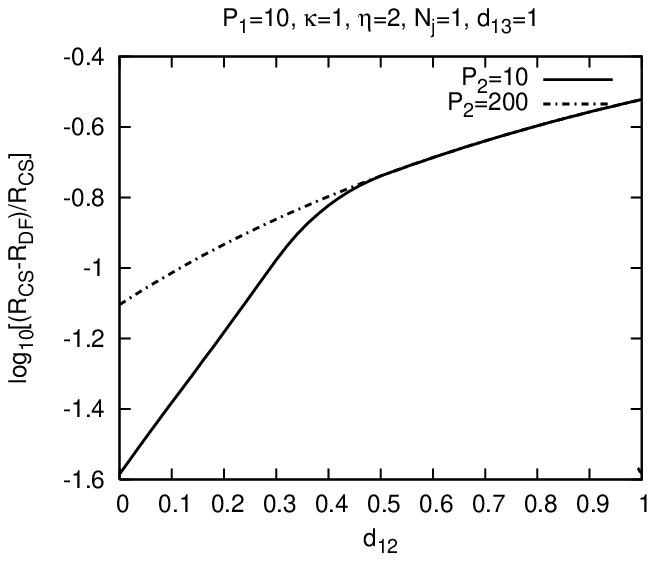}
\caption{The performance of the decode-and-forward strategy in the MPL model.}
\label{fig:relay_close_to_source_vary_distance_mpl}
\end{minipage}
\end{figure}

\begin{figure}[t]
\centering
\includegraphics[width=0.5\textwidth]{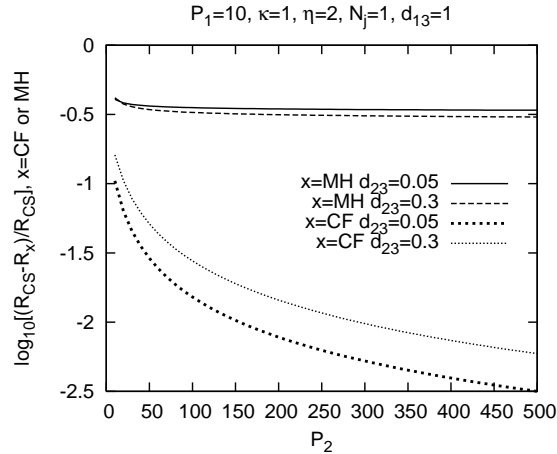}
\caption{The performance of the compress-and-forward strategy and the TDMA multihop strategy in the MPL model.}
\label{fig:relay_close_to_destination_vary_power_mpl}
\end{figure}

Table~\ref{tab:relay_distance_power_mpl} summarizes the results for the MPL model. In the MPL model, $\lambda_{12}$ and $\lambda_{23}$ do not approach infinity even if the source-relay distance or the relay-destination distance approaches zero. Here, we see that when the relay power is limited, none of the strategies here achieves the capacity regardless of the relay position. However, when the relay power grows, the compress-and-forward strategy achieves the capacity. The proof of the results can be found in Appendix~\ref{append:relay_distance_power_mpl}. The results of numerical computations can be found in Figs.~\ref{fig:relay_close_to_destination_vary_distance_mpl}--\ref{fig:relay_close_to_destination_vary_power_mpl}. In Figs.~\ref{fig:relay_close_to_destination_vary_distance_mpl} and \ref{fig:relay_close_to_source_vary_distance_mpl}, we show that the compress-and-forward strategy does not achieve the capacity in the MPL model even when the relay-destination distance goes to zero, and the decode-and-forward strategy does not achieve the capacity in the MPL model even when the relay-source distance goes to zero.

However, we see in Fig.~\ref{fig:relay_close_to_destination_vary_power_mpl} that the rate achievable by the compress-and-forward strategy approaches the capacity as the relay power grows. In addition, the graph shows that it is not necessary for the relay to transmit at infinitely large power for the rate of the compress-and-forward strategy to get near the capacity. At $d_{23} = 0.05$, the compress-and-forward rate gets to 97\% of the capacity when the relay transmits at power five times the source power; at $d_{23}=0.3$, the compress-and-forward rate gets to 97\% of the capacity when the relay transmits at power nine times the source power. In the same figure, we also find that the performance of the TDMA multihop strategy is far below that of the compress-and-forward strategy.

\section{Conclusions} \label{sec:mesh_conclusions}
In this paper, we studied the SSMRSD mesh network. We showed that the compress-and-forward strategy achieves the capacity as the relay powers get asymptotically large. We performed numerical computations on the single relay channel using two signal propagation models, i.e, the SPL model and the MPL model. In the SPL model, the compress-and-forward strategy approaches the capacity when the relay is close to the destination, and the decode-and-forward strategy approaches the capacity when the relay is close to the source. These happen because the channel gains approach infinity as the inter-node distances go to zero.  The MPL model rectifies this anomalous behavior. In the MPL model, the compress-and-forward strategy only approaches the capacity when the relay power gets large. In other words, none of the strategies achieves the capacity with low relay power regardless of the relay position.  We observe (in the single relay channel) that the compress-and-forward strategy can get close to the capacity when the relay power is just several times larger than the source power.  We also note that the compress-and-forward strategy does better than the TDMA multihop strategy.
 
%We show that the performance of the compress-and-forward strategy gets close to the capacity when the relay's power is a few times higher than the source power, meaning that we do not need infinite relay's power to get close to the capacity.

%A multiple source multiple destination mesh network can be viewed as a collection of many SSMRSD mesh networks.

For SSMRSD mesh network design, our results suggest that information theoretic cooperative relaying (via the compress-and-forward strategy) by mesh routers is a good alternative to conventional multihop  communication between mesh clients.
Recent work \cite{huli06} indicates how one can design practical coding schemes to implement cooperative relaying.
Our results can be used to determine what power the mesh routers should transmit at, and how the nodes should cooperate. 

The relevance of the SSMRSD network is in a multiple source multiple destination network in which the channels of all source-destination pairs (including their relays) are orthogonal. In this situation, each source-destination pair, together with their assigned relays, operates as an SSMRSD network. 
Combining this with network layer algorithms can lead to an efficient protocol stack for mesh networks, on which a variety of user applications can be built.

%\newpage
\bibliography{bib}

\appendix

\section{Proof of Table~\ref{tab:relay_distance_power}} \label{append:relay_distance_power}
\textbf{Case~\eqref{lab:case4}:} $P_2 \rightarrow \infty, d_{12} \rightarrow 0, \frac{P_2}{d_{12}^{-\eta}} \rightarrow \infty$ or equivalently $\lambda_{12} \rightarrow \infty, \frac{P_2}{\lambda_{12}} \rightarrow \infty$.\\
Setting $\alpha_{\text{CS}} =0$ maximizes the cut-set rate.
\begin{subequations}
\begin{align}
R_{\text{CS}} &= L\left( \frac{P_1\lambda_{13}}{N_3} + \frac{P_1 \lambda_{12}}{N_2} \right) \\
& \rightarrow L\left( \frac{P_1 \lambda_{12}}{N_2} \right) \quad \text{as $d_{12}\rightarrow 0$}.
\end{align}
\end{subequations}

Using the decode-and-forward strategy, we set $\alpha_{\text{DF}}=0$. The rate achievable is
\begin{subequations}
\begin{align}
R_{\text{DF}} & = L \left( \frac{P_1 \lambda_{12}}{N_2} \right)\\
R_{\text{DF}} & \rightarrow R_{\text{CS}} \quad \text{as $d_{12}$ decreases} \\
R_{\text{DF}} & \approx R_{\text{CS}} \quad \text{as $P_2$ increases}.
\end{align}
\end{subequations}
So, for case~\eqref{lab:case4}, the decode-and-forward strategy approaches the capacity as the distance $d_{12}$ goes smaller. It does not approaches the capacity, but stay close to the capacity, when the relay power increases.

Using the compress-and-forward strategy, noting that $\frac{\lambda_{12}}{P_2} \rightarrow 0$,
\begin{subequations}
\begin{align}
R_{\text{CF}} & = L\left( \frac{P_1\lambda_{13}}{N_3} + \frac{P_1P_2\lambda_{12}\lambda_{23}}{P_1(\lambda_{13}N_2 + \lambda_{12}N_3) +P_2\lambda_{23}N_2 + N_2N_3}  \right) \\
& = L\left( \frac{P_1\lambda_{13}}{N_3} + \frac{P_1\lambda_{12}}{N_2 + \frac{1}{P_2}\frac{P_1\lambda_{13}N_2}{\lambda_{23}} + \frac{\lambda_{12}}{P_2}\frac{P_1N_3}{\lambda_{23}} + \frac{1}{P_2}\frac{N_2N_3}{\lambda_{23}} } \right) \\
& \rightarrow R_{\text{CS}} \quad \text{as either $P_2$ increases or $d_{12}$ decreases}.
\end{align}
\end{subequations}

We note that the while only decreasing $d_{12}$ drives rate of the decode-and-forward rate to the capacity, increasing $P_2$ or decreasing $d_{12}$ can drive the compress-and-forward rate to the capacity.

\textbf{Case \eqref{lab:case5}:} $P_2 \rightarrow \infty, d_{12} \rightarrow 0, \frac{P_2}{d_{12}^{-\eta}} \approx 1$ or equivalently $\lambda_{12} \rightarrow \infty, \frac{P_2}{\lambda_{12}} \approx 1$.\\
Under this condition, the cut-set bound is given $R_{\text{CS}}$ that satisfies the conditions below, for some $0 < \alpha_{\text{CS}} < 1$.
\begin{subequations}
\begin{align}
R_{\text{CS}} & = L\left( \frac{P_1\lambda_{13}}{N_3} + \frac{P_2\lambda_{23}}{N_3} + \frac{2 \sqrt{\alpha_{\text{CS}}  \lambda_{13} \lambda_{23} P_1 P_2 }}{N_3}    \right) \\
& = L\left( \frac{P_2\lambda_{23}}{N_3} + \frac{2 \sqrt{\alpha_{\text{CS}}  \lambda_{13} \lambda_{23} P_1 P_2 }}{N_3}  \right)^+, \\
R_{\text{CS}} & = L\left( \left(\frac{P_1\lambda_{13}}{N_3} + \frac{P_1 \lambda_{12}}{N_2} \right) (1 - \alpha_{\text{CS}}) \right) \\
& = L\left( \frac{P_1 \lambda_{12}}{N_2} (1 - \alpha_{\text{CS}}) \right)^+,
\end{align}
\end{subequations}

The decode-and-forward strategy achieves the rate up to $R_{\text{DF}}$ that satisfies the following conditions, for some $0 < \alpha_{\text{DF}} < \alpha_{\text{CS}}$. We note that $\alpha_{\text{DF}} \approx \alpha_{\text{CS}}$.
\begin{subequations}
\begin{align}
R_{\text{DF}} & = L\left( \frac{P_1\lambda_{13}}{N_3} + \frac{P_2\lambda_{23}}{N_3} + \frac{2 \sqrt{\alpha_{\text{DF}}  \lambda_{13} \lambda_{23} P_1 P_2 }}{N_3} \right) \\
& = L\left( \frac{P_2\lambda_{23}}{N_3} +  \frac{2 \sqrt{\alpha_{\text{DF}}  \lambda_{13} \lambda_{23} P_1 P_2 }}{N_3} \right)^+ \\
& \nrightarrow R_{\text{CS}} \quad \text{as $P_2$ increases because $\alpha_{\text{DF}}P_2 << \alpha_{\text{CS}}P_2$}, \\
R_{\text{DF}} & = L\left( \frac{P_1 \lambda_{12}}{N_2} (1 - \alpha_{\text{DF}}) \right) \\
& \rightarrow R_{\text{CS}} \quad \text{as $d_{12}$ decreases because $(1-\alpha_{\text{DF}}) \approx (1-\alpha_{\text{CS}})$},
\end{align}
\end{subequations}

Using the compress-and-forward strategy, the following rate is achievable.
\begin{subequations}
\begin{align}
R_{\text{CF}} & = L\left( \frac{P_1\lambda_{13}}{N_3} + \frac{P_1P_2\lambda_{12}\lambda_{23}}{P_1(\lambda_{13}N_2 + \lambda_{12}N_3) +P_2\lambda_{23}N_2 + N_2N_3}  \right) \\
& = L\left( \frac{P_1\lambda_{13}}{N_3} + \frac{P_1\lambda_{12}}{N_2 + \frac{1}{P_2}\frac{P_1\lambda_{13}N_2}{\lambda_{23}} + \frac{\lambda_{12}}{P_2}\frac{P_1N_3}{\lambda_{23}} + \frac{1}{P_2}\frac{N_2N_3}{\lambda_{23}} } \right) \\
& \approx L\left( \frac{P_1\lambda_{12}}{N_2 + \frac{\lambda_{12}}{P_2}\frac{P_1N_3}{\lambda_{23}}  } \right).
\end{align}
\end{subequations}
It is not clear if $R_{\text{CF}}$ approaches $R_{\text{CS}}$.

\textbf{Case \eqref{lab:case6}:} $P_2 \rightarrow \infty, d_{12} \rightarrow 0, \frac{P_2}{d_{12}^{-\eta}} \rightarrow 0$ or equivalently $\lambda_{12} \rightarrow \infty, \frac{P_2}{\lambda_{12}} \rightarrow 0$.\\
The optimal power splits are $\alpha_{\text{CS}} \rightarrow 1$ and $\alpha_{\text{DF}} \rightarrow 1$ but $\alpha_{\text{DF}} < \alpha_{\text{CS}}$. Using the same reasoning as that in case~\eqref{lab:case5},
\begin{subequations}
\begin{align}
R_{\text{DF}} & \nrightarrow R_{\text{CS}} \quad \text{, as $P_2$ increases because $\alpha_{\text{DF}}P_2 << \alpha_{\text{CS}}P_2$}, \\
R_{\text{DF}} & \rightarrow R_{\text{CS}} \quad \text{, as $d_{12}$ decreases because $(1-\alpha_{\text{DF}}) \approx (1-\alpha_{\text{CS}})$}.
\end{align}
\end{subequations}

The cut set bound is
\begin{subequations}
\begin{align}
R_{\text{CS}} & = L\left( \frac{P_1\lambda_{13}}{N_3} + \frac{P_2\lambda_{23}}{N_3} + \frac{2 \sqrt{\alpha_{\text{CS}}  \lambda_{13} \lambda_{23} P_1 P_2 }}{N_3}    \right) \\
& = L\left( \frac{P_2\lambda_{23}}{N_3} + \frac{2 \sqrt{\alpha_{\text{CS}}  \lambda_{13} \lambda_{23} P_1 P_2 }}{N_3}  \right)^+.
\end{align}
\end{subequations}
Note that $\alpha_{\text{CS}}$ increases as $d_{12}$ decreases.

The compress-and-forward strategy achieves the following rate.
\begin{subequations}
\begin{align}
R_{\text{CF}} & = L\left( \frac{P_1\lambda_{13}}{N_3} + \frac{P_1 P_2 \lambda_{12} \lambda_{23} } {P_1(\lambda_{13}N_2 + \lambda_{12}N_3) +P_2\lambda_{23}N_2 + N_2N_3}  \right) \\
& = L\left( \frac{P_1\lambda_{13}}{N_3} + \frac{P_1P_2\lambda_{23}}{P_1N_3 + \frac{P_2}{\lambda_{12}}\lambda_{23}N_2 + \frac{1}{\lambda_{12}}( P_1 \lambda_{13} N_2 + N_2N_3 ) } \right), \\
& = L\left( \frac{P_2\lambda_{23}}{N_3} \right)^+ \\
R_{\text{CF}} & \nrightarrow R_{\text{CS}} \quad \text{as $d_{12}$ decreases because $\alpha_{\text{CS}}$ increases,} \\
R_{\text{CF}} & \rightarrow R_{\text{CS}} \quad \text{as $P_2$ increases because $P_2$ increases faster than $\sqrt{P_2}$.}
\end{align}
\end{subequations}

\textbf{Cases \eqref{lab:case7} and \eqref{lab:case8}:} Relay power $P_2 \rightarrow \infty$. Source-relay distance $d_{12} = K_2$ and $d_{23} \rightarrow 0$ for cases~\eqref{lab:case7} and \eqref{lab:case8} respectively.\\
The cut-set bound is
\begin{equation}
R_{\text{CS}} = L\left( \frac{P_1\lambda_{13}}{N_3} + \frac{P_1 \lambda_{12}}{N_2}  \right).
\end{equation}
The decode-and-forward strategy is optimized by setting $\alpha_{\text{DF}} = 0$. It achieves the following rate.
\begin{equation}
R_{\text{DF}} = L\left( \frac{P_1 \lambda_{12}}{N_2}  \right) < R_{\text{CS}}.
\end{equation}
The compress-and-forward strategy achieves the rate up to
\begin{subequations}
\begin{align}
R_{\text{CF}} & = L\left( \frac{P_1\lambda_{13}}{N_3} + \frac{P_1 P_2 \lambda_{12} \lambda_{23} } {P_1(\lambda_{13}N_2 + \lambda_{12}N_3) +P_2\lambda_{23}N_2 + N_2N_3}  \right) \\
& = L\left( \frac{P_1\lambda_{13}}{N_3} + \frac{P_1\lambda_{12}}{N_2 + \frac{1}{P_2\lambda_{23}}(P_1\lambda_{13}N_2 + P_1\lambda_{12}N_3 + N_2N_3) }\right) \\
& = L\left( \frac{P_1\lambda_{13}}{N_3} + \frac{P_1 \lambda_{12}}{N_2}  \right)^- \quad \text{as $P_2$ increases or $d_{23}$ decreases} \\
& \rightarrow R_{\text{CS}} \quad \text{as $P_2$ increases or $d_{23}$ decreases}.
\end{align}
\end{subequations}

\textbf{Case \eqref{lab:case1}:} $P_2 = K_1, d_{12} \rightarrow 0$ or $\lambda_{12} =\kappa d_{12}^{-\eta} \rightarrow \infty$.\\
The cut-set bound is obtained by setting $\alpha_{\text{CS}}$ such that
\begin{equation}
\left(\frac{P_1\lambda_{13}}{N_3} + \frac{P_1 \lambda_{12}}{N_2} \right) (1 - \alpha_{\text{CS}})  =  \frac{P_1\lambda_{13}}{N_3} + \frac{P_2\lambda_{23}}{N_3} + \frac{2 \sqrt{\alpha_{\text{CS}}  \lambda_{13} \lambda_{23} P_1 P_2 }}{N_3}.
\end{equation}
The cut set bound is thus
\begin{equation}
R_{\text{CS}} = L\left( \frac{P_1\lambda_{13}}{N_3} + \frac{P_2\lambda_{23}}{N_3} + \frac{2 \sqrt{ \alpha_{\text{CS}} \lambda_{13} \lambda_{23} P_1 P_2 }}{N_3}    \right),
\end{equation}
for some $\alpha_{\text{CS}} \rightarrow 1$ as $d_{12} \rightarrow 0$.

Using the decode-and-forward strategy, we choose $\alpha_{\text{DF}}$ such that
\begin{equation}
 \frac{P_1 \lambda_{12}}{N_2} (1 - \alpha_{\text{DF}})  =  \frac{P_1\lambda_{13}}{N_3} + \frac{P_2\lambda_{23}}{N_3} + \frac{2 \sqrt{\alpha_{\text{DF}}  \lambda_{13} \lambda_{23} P_1 P_2 }}{N_3}.
\end{equation}
Note that $\alpha_{\text{DF}} \rightarrow 1$ but $\alpha_{\text{DF}} < \alpha_{\text{CS}}$. As $d_{12} \rightarrow 0$, $\alpha_{\text{DF}} \rightarrow \alpha_{\text{CS}}$. Hence, we get
\begin{subequations}
\begin{align}
R_{\text{DF}} & = L\left( \frac{P_1\lambda_{13}}{N_3} + \frac{P_2\lambda_{23}}{N_3} + \frac{2 \sqrt{ \alpha_{\text{DF}} \lambda_{13} \lambda_{23} P_1 P_2 }}{N_3}    \right) \\
& \rightarrow R_{\text{CS}} \quad \text{as $d_{12} \rightarrow 0$}.
\end{align}
\end{subequations}

Using the compress-and-forward strategy,
\begin{subequations}
\begin{align}
Q & = \frac{(\lambda_{13}N_2 + \lambda_{12}N_3)P_1 + N_2N_3}{P_2 \lambda_{23}} \rightarrow \infty \\
R_{\text{CF}} & = L\left( \frac{P_1 \lambda_{13}}{N_3} \right) < R_{\text{CS}}.
\end{align}
\end{subequations}

\textbf{Case \eqref{lab:case2}:} $P_2 = K_1, d_{12} =K_2$ or $\lambda_{12} = K_4$ for some finite $K_4$.\\
The cut-set bound is
\begin{subequations}
\begin{align}
R_{\text{CS}} & = L\left( \left(\frac{P_1\lambda_{13}}{N_3} + \frac{P_1 \lambda_{12}}{N_2} \right) (1 - \alpha_{\text{CS}}) \right) \label{eq:set-set_1} \\
& = L\left( \frac{P_1\lambda_{13}}{N_3} + \frac{P_2\lambda_{23}}{N_3} + \frac{2 \sqrt{\alpha_{\text{CS}}  \lambda_{13} \lambda_{23} P_1 P_2 }}{N_3}    \right), \label{eq:set-set_2}
\end{align}
\end{subequations}
for some $\alpha_{\text{CS}}$ or
\begin{equation}
R_{\text{CS}} = L\left( \frac{P_1\lambda_{13}}{N_3} + \frac{P_1 \lambda_{12}}{N_2} \right),
\end{equation}
if the optimal $\alpha_{\text{CS}} = 0$.

Using the decode-and-forward strategy,
\begin{subequations}
\begin{align}
R_{\text{DF}} & = L\left(  \frac{P_1 \lambda_{12}}{N_2} (1 - \alpha_{\text{DF}}) \right) \\
& = L\left( \frac{P_1\lambda_{13}}{N_3} + \frac{P_2\lambda_{23}}{N_3} + \frac{2 \sqrt{\alpha_{\text{DF}}  \lambda_{13} \lambda_{23} P_1 P_2 }}{N_3}    \right) \\
& < R_{\text{CS}},
\end{align}
\end{subequations}
for some $\alpha_{\text{DF}} < \alpha_{\text{CS}}$ or
\begin{equation}
R_{\text{DF}} = L\left(  \frac{P_1 \lambda_{12}}{N_2} \right) < R_{\text{CS}},
\end{equation}
if the optimal $\alpha_{\text{DF}} = 0$.

Using the compress-and-forward strategy,
\begin{equation}
R_{\text{CF}} = L\left( \frac{P_1\lambda_{13}}{N_3} + \frac{P_1P_2\lambda_{12}\lambda_{23}}{P_1(\lambda_{13}N_2 + \lambda_{12}N_3) +P_2\lambda_{23}N_2 + N_2N_3}  \right).
\end{equation}
It is not clear how the compress-and-forward strategy performs compared to the cut-set bound. However, an example \cite{kramergastpar04} shows that it can be ``far'' below the cut-set bound.

\textbf{Case \eqref{lab:case3}:} $P_2 = K_1, d_{23} \rightarrow 0$ or $\lambda_{23} = \kappa d_{23}^{-\eta} \rightarrow \infty$.\\
Under this condition, setting $\alpha_{\text{CS}} = 0$ maximizes the cut-set rate.
\begin{equation}
R_{\text{CS}} = L\left( \frac{P_1\lambda_{13}}{N_3} + \frac{P_1 \lambda_{12}}{N_2} \right).
\end{equation}
Setting $\alpha_{\text{DF}}=0$ maximizes the achievability of the decode-and-forward strategy.
\begin{equation}
R_{\text{DF}} = L\left( \frac{P_1\lambda_{13}}{N_3}  \right) < R_{\text{CS}}.
\end{equation}
Using the compress-and-forward strategy,
\begin{subequations}
\begin{align}
R_{\text{CF}} & = L\left( \frac{P_1\lambda_{13}}{N_3} + \frac{P_1 P_2 \lambda_{12} \lambda_{23} } {P_1(\lambda_{13}N_2 + \lambda_{12}N_3) +P_2\lambda_{23}N_2 + N_2N_3}  \right) \\
& = L\left( \frac{P_1\lambda_{13}}{N_3} + \frac{P_1\lambda_{12}}{N_2 + \frac{1}{\lambda_{23}}\left(\frac{P_1\lambda_{13}N_2 + P_1\lambda_{12}N_3 + N_2N_3}{P_2}\right) }\right) \\
& \rightarrow R_{\text{CS}} \quad \text{as $d_{23} \rightarrow 0$}.
\end{align}
\end{subequations}

\section{Proof of Table~\ref{tab:relay_distance_power_mpl}} \label{append:relay_distance_power_mpl}
\textbf{Case~\eqref{lab:case9}:} The proof is similar to that for case~\eqref{lab:case7}. The compress-and-forward strategy achieves the capacity but not the decode-and-forward strategy.

\textbf{Case~\eqref{lab:case10}:} The proof is similar to that for case~\eqref{lab:case2}, in which $P_2, \lambda_{12},$ and $\lambda_{23}$ are finite. Neither the decode-and-forward strategy nor the compress-and-forward strategy achieves the capacity.

\end{document}